\documentclass[10pt, final, journal, letterpaper, oneside, twocolumn]{IEEEtran}
\usepackage{graphicx}
\usepackage{amsmath}
\usepackage{amssymb}
\usepackage{mathrsfs}
\usepackage{stfloats}
\usepackage{dsfont}
\usepackage{setspace}
\usepackage{array}
\usepackage{bbm}
\usepackage{caption}
\usepackage{epstopdf}
\usepackage{xcolor}
\usepackage{accents}

\newtheorem{definition}{\bf{Definition}}
\newtheorem{theorem}{\bf{Theorem}}
\newtheorem{corollary}{\bf{Corollary}}

\newtheorem{lemma}{\bf{Lemma}}

\DeclareMathAlphabet\mathbfcal{OMS}{cmsy}{b}{n}
\begin{document}
\title{\Large End-to-End Mutual Coupling Aware Communication Model for Reconfigurable Intelligent Surfaces: An Electromagnetic-Compliant Approach Based on Mutual Impedances}
\author{Gabriele~Gradoni, \IEEEmembership{Member,~IEEE} and Marco~Di~Renzo, \IEEEmembership{Fellow,~IEEE} \vspace{-0.75cm}
\thanks{Manuscript received Sep. 6, 2020; revised Dec. 7, 2020. G. Gradoni is with BT and Univ. of Nottingham, UK. M. Di Renzo is with CNRS and Paris-Saclay Univ., France (e-mail: marco.di-renzo@universite-paris-saclay.fr).}
}
%
%
%
%
\maketitle
\begin{abstract}
Reconfigurable intelligent surfaces (RISs) are an emerging technology for application to wireless networks. We introduce a physics and electromagnetic (EM) compliant communication model for analyzing and optimizing RIS-assisted wireless systems. The proposed model has four main notable attributes: (i) it is \textit{end-to-end}, i.e., it is formulated in terms of an equivalent channel that yields a one-to-one mapping between the voltages fed into the ports of a transmitter and the voltages measured at the ports of a receiver; (ii) it is \textit{EM-compliant}, i.e., it accounts for the generation and propagation of the EM fields; (iii) it is \textit{mutual coupling aware}, i.e., it accounts for the mutual coupling among the sub-wavelength unit cells of the RIS; and (iv) it is \textit{unit cell aware}, i.e., it accounts for the intertwinement between the amplitude and phase response of the unit cells of the RIS.
\end{abstract}
\vspace{-0.17cm}
\begin{IEEEkeywords}
Wireless, reconfigurable intelligent surfaces.
\end{IEEEkeywords}
%
%
%
%
\vspace{-0.35cm}
\section{Introduction} \label{Introduction} \vspace{-0.1cm}
Reconfigurable intelligent surfaces (RISs) are an emerging and promising software-defined technology for enhancing the performance of wireless networks at a low cost, power, and complexity \cite{MDR_Eurasip}, \cite{MDR_Relays}. An RIS consists of a large number of inexpensive and nearly-passive scattering elements that can be configured to customize the propagation of the radio waves \cite{MDR_JSAC}.

For analyzing and optimizing RIS-aided wireless systems, sufficiently realistic and accurate yet tractable communication models that account for the physics and electromagnetic (EM) properties of the scattering elements of the RIS are needed. This is an open research issue, and only a few EM-compliant models for RIS-assisted wireless systems are available to date \cite{Wankai}-\cite{Marzetta}. In \cite{Wankai}, an experimentally-validated path-loss model for a non-homogenizable RIS is proposed. In \cite{MDR_PathLoss}, a path-loss model for a homogenizable RIS is introduced by using the vector theory of scattering. In \cite{RuiZhang}, an approach for modeling the interplay between the amplitude and phase response of a lossy unit cell is proposed. In \cite{Marzetta}, the mutual coupling among the active radiating elements of a large surface is investigated.

These research works are noteworthy but tackle specific issues in isolation, and do not provide an operational and end-to-end communication model for analyzing and optimizing RIS-assisted wireless systems, in which the impact of the EM fields and of the currents impressed and induced on the radiating elements can be explicitly identified. In this letter, we tackle these fundamental issues by introducing a new physics- and EM-compliant communication model for RIS-assisted wireless systems. The proposed approach has four main attributes: (i) it has its foundation on the laws of electromagnetism for the generation, propagation, and scattering of the EM fields; (ii) it accounts for the intertwinement between the amplitude and phase response and for the mutual coupling between the radiated EM fields and the current induced on closely spaced scatterers; and (iii) it yields a one-to-one mapping between the voltages measured at the ports of a multi-antenna receiver and the voltages impressed at the ports of a multi-antenna transmitter. Thus, the proposed communication model is end-to-end, EM-compliant, mutual coupling and unit cell aware.

More specifically, we introduce a circuit-based communication model for RIS-assisted wireless systems that is based on the mutual impedances between all existing radiating elements (transmit/receive antennas, passive scatterers) \cite{Gradoni}. We prove that the impact of (i) the incident EM fields (impressed by a generator or scattered/induced); (ii) the voltage generators at the transmitter and the load impedances at the receiver; and (iii) the tunable load impedances that control the passive scatterers of the RIS is jointly taken into account, and is explicitly unfolded and individually observable in the proposed end-to-end communication model. To the best of our knowledge, this letter introduces the first complete EM-compliant communication model for RIS-aided wireless systems.

\textit{Notation}: Vectors and matrices are denoted in bold font; ${{\bf{\hat z}}}$ is the unit-norm vector $\bf{z}$; $j=\sqrt{-1}$ is the imaginary unit; $\varepsilon_0$ and $\mu_0$ are the permittivity and permeability in vacuum; ${c_0} = 1/\sqrt {{\varepsilon _0}{\mu _0}}$ is the speed of light; $\eta_0 = \sqrt{\mu_0/\varepsilon_0}$ is the characteristic independence in vacuum; $f$ is the frequency; $\omega  = 2\pi f$ is the angular frequency; $\lambda$ is the wavelength; $k_0=2 \pi /\lambda$ is the wavenumber; $\delta(\cdot)$ is the Dirac delta function; ${\partial _z^2}$ denotes the second-order partial derivative with respect to $z$; and $\cdot$ denotes the scalar product between vectors.

\begin{figure}[!t]
\centering
\includegraphics[width=\columnwidth]{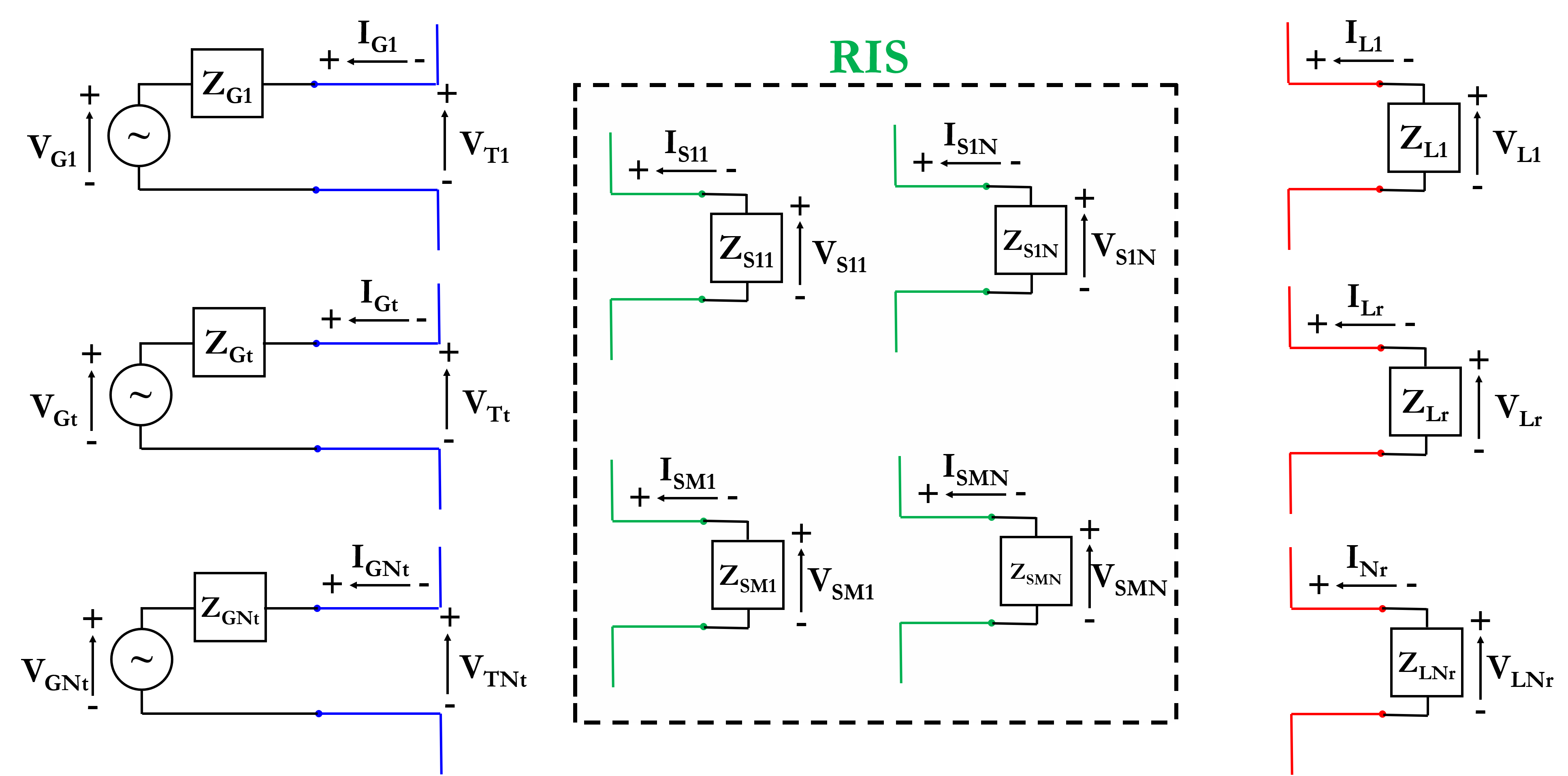}
\vspace{-0.6cm} \caption{System-model.}
\label{Fig_1} \vspace{-0.5cm}
\end{figure}
\vspace{-0.35cm}
\section{System Model} \label{System Model} \vspace{-0.1cm}
We consider the wireless system in Fig. \ref{Fig_1}, in which a transmitter with $N_{\rm{t}}$ antennas and a receiver with $N_{\rm{r}}$ antennas communicate through an RIS that is made of $N_{\rm{ris}} = M \times N$ passive scatterers. The locations of the $t$th transmit antenna, $r$th receive antenna, and $s$th passive scatterer are ${{\bf{r}}_{\xi}} = {x_{\xi}}{\bf{\hat x}} + {y_{\xi}}{\bf{\hat y}} + {z_{\xi}}{\bf{\hat z}}$ for ${\xi} = \{ t,r,s\}$. For illustrative purposes, the transmit antennas, the receive antennas, and the passive scatterers are cylindrical thin wires of perfectly conducting material whose length is $l$ and whose radius $a \ll l$ is finite but negligible with respect to $l$ (thin wire regime). The inter-distance between adjacent radiating elements is denoted by $d$. For generality, we consider three distinct triplets $(l_{\xi}, a_{\xi}, d_{\xi})$. All thin wires are parallel to the ${\bf{\hat z}}$-axis. The proposed model is general enough for application to different types of antennas, passive scatterers, spatial distributions, provided that the radiating elements are minimum scattering antennas \cite{Gradoni}.

\vspace{-0.35cm}
\subsection{Transmitter Modeling} \label{Transmitter} \vspace{-0.10cm}
Each transmit antenna is fed by an independent voltage generator $V_{\rm{Gt}}$, $t=1,2,\ldots, N_{\rm{t}}$, according to the delta-gap model \cite[Eq. 8.28]{Balanis} that imitates an antenna fed by a transmission line. As shown in Fig. \ref{Fig_1}, the generator is applied between the lower and upper halves of the transmit antenna across a short gap of negligible size. The voltage generator has an internal impedance $Z_{\rm{Gt}}$, and, therefore, the current flowing through the port of the transmit antenna is $I_{\rm{Tt}} = (V_{\rm{Gt}} - V_{\rm{Tt}})/Z_{\rm{Gt}}$, where $V_{\rm{Tt}}$ is the voltage at the input of the port. In our system model, $V_{\rm{Gt}}$ and $Z_{\rm{Gt}}$ are given, and $I_{\rm{Tt}}$ and $V_{\rm{Tt}}$ are variables to be determined as elaborated next.

\vspace{-0.35cm}
\subsection{Receiver Modeling} \label{Receiver} \vspace{-0.10cm}
Each receive antenna is connected to an independent load impedance $Z_{\rm{Lr}}$, $r=1,2,\ldots, N_{\rm{r}}$. Therefore, the relation between the voltage drop across the load, $V_{\rm{Lr}}$, and the current flowing through the load, $I_{\rm{Lr}}$, is $I_{\rm{Lr}} = - V_{\rm{Lr}}/Z_{\rm{Lr}}$. Since the load impedance is assumed to be infinitesimally small, this model is usually referred to as the impedance delta-gap model. In our system model, $Z_{\rm{Lr}}$ is given, and $I_{\rm{Lr}}$ and $V_{\rm{Lr}}$ are variables to be determined as elaborated next.

\vspace{-0.35cm}
\subsection{RIS Modeling} \label{RIS} \vspace{-0.10cm}
The RIS is modeled as a collection of $N_{\rm{ris}}$ passive scatterers. Each passive scatterer is a cylindrical thin wire of perfectly conducting material, which is connected, between the lower and upper halves of the wire, to a tunable load $Z_{\rm{Smn}}$, $m=1,2,\ldots, M$ and $n=1,2,\ldots, N$. Similar to the receiver, the relation between the voltage drop across the load, $V_{\rm{Smn}}$, and the current flowing through the load, $I_{\rm{Smn}}$, is $I_{\rm{Smn}} = - V_{\rm{Smn}}/Z_{\rm{Smn}}$. Also, $Z_{\rm{Smn}}$ is given, and $I_{\rm{Smn}}$ and $V_{\rm{Smn}}$ are variables to be determined as elaborated next.

By appropriately optimizing the tunable loads $Z_{\rm{Smn}}$, one can adaptively configure the scattered EM fields. Assuming, e.g., that each tunable load is a positive intrinsic negative (PIN) diode, the corresponding impedance is ${Z_{{\rm{Smn}}}} = {R_{{\rm{Smn}}}} + j\omega {L_{{\rm{Smn}}}}$ and ${Z_{{\rm{Smn}}}} = Z_{{\rm{Smn}}}^\parallel  + j\omega {L_{{\rm{Smn}}}}$ with ${1 \mathord{\left/ {\vphantom {1 {Z_{{\rm{Smn}}}^\parallel }}} \right. \kern-\nulldelimiterspace} {Z_{{\rm{Smn}}}^\parallel }} = {1 \mathord{\left/ {\vphantom {1 {{R_{{\rm{Smn}}}}}}} \right. \kern-\nulldelimiterspace} {{R_{{\rm{Smn}}}}}} + j\omega {C_{{\rm{Smn}}}}$ for forward and negative bias implementations, respectively, where ${R_{{\rm{Smn}}}}$, ${L_{{\rm{Smn}}}}$, and ${C_{{\rm{Smn}}}}$ are the resistance, inductance, and capacitance of the PIN diode. By varying the inductance and/or the capacitance, one can reconfigure the response of each passive scatterer. The resistance usually accounts for the internal losses of the PIN diode.

\vspace{-0.35cm}
\subsection{Modeling Assumptions and Methodology} \label{Assumptions} \vspace{-0.10cm}
Our objective is to introduce a physics- and EM-compliant communication model for RIS-assisted communications that is accurate but tractable and insightful enough for application to wireless systems. Due to the complexity of the generation, propagation, and scattering of the EM waves, some assumptions are necessary. In this section, we clarify the assumptions of our system formulation, by focusing on the transmission, interaction, and detection of the EM waves in free space. The generalization to random media is postponed to a future work.

An antenna, whether transmitting or receiving, is driven by an external electric field that acts as a source. Based on Section \ref{Transmitter}, a transmit antenna is driven by a voltage generator that is applied to its port. Thus, the source electric field is determined by the voltage generator. A receive antenna, e.g., for the receiver and the RIS in Sections \ref{Receiver} and \ref{RIS}, respectively, is driven by the external electric field (generated by any physical source, e.g., a transmit antenna driven by a voltage generator) that impinges upon it. In both cases, the source electric field is referred to as the \textit{incident} field and is denoted by ${\bf{E}}^{(\rm{inc})}$. This incident field induces a current on the antenna, either in transmit or receive mode, and the induced current generates, in turn, its own electric field that is radiated away from the antenna. This electric field is referred to as the \textit{radiated} field and is denoted by ${\bf{E}}^{(\rm{rad})}$. For any antenna, thus, the electric field at any observation point, $\bf{r}$, is the sum of the incident and radiated fields, i.e., ${\bf{E}}(\bf{r})= {\bf{E}}^{(\rm{inc})}(\bf{r}) + {\bf{E}}^{(\rm{rad})}(\bf{r})$. 

Given ${\bf{E}}^{(\rm{inc})}$, the key component to develop an EM-compliant communication model lies in determining the distribution of the current induced on (the surface of) each radiating element. This current distribution is determined by the boundary conditions of the electric field on the surface of each antenna. When the radiating elements are made of perfectly conducting material, the boundary conditions impose that the tangential component of the electric field vanishes on the surface of the antenna. In Fig. \ref{Fig_1}, the thin wire antennas are parallel to the ${\bf{\hat z}}$-axis. Thus, the tangential component of the electric field is ${E_z}({\bf{r}}) = {\bf{E}}({\bf{r}}) \cdot {\bf{\hat z}} = {{\bf{E}}^{\left( {{\rm{inc}}} \right)}}({\bf{r}}) \cdot {\bf{\hat z}} + {{\bf{E}}^{\left( {{\rm{rad}}} \right)}}({\bf{r}}) \cdot {\bf{\hat z}} = E_z^{\left( {{\rm{inc}}} \right)}({\bf{r}}) + E_z^{\left( {{\rm{rad}}} \right)}({\bf{r}})$, and the boundary conditions at any point on the surface, ${\bf r}_\mathcal{S}$, of any (transmit or receive) radiating element impose ${E_z}\left( {{\bf r}_\mathcal{S}} \right) = 0$, i.e., $E_z^{\left( {{\rm{rad}}} \right)}\left( {{\bf r}_\mathcal{S}} \right) =  - E_z^{\left( {{\rm{inc}}} \right)}\left( {{\bf r}_\mathcal{S}} \right)$.

Consider a generic radiating element located at ${{\bf{r}}_{\xi}}$. The tangential component of the incident field on its surface, $E_z^{\left( {{\rm{inc}}} \right)}\left( {{\bf r}_\mathcal{S}} \right)$, is, thus, sufficient for computing the distribution of the current. Let ${\bf{I}}\left( {{z^{\prime}}} \right)$, for $ z_{\xi} - {l \mathord{\left/ {\vphantom {l 2}} \right. \kern-\nulldelimiterspace} 2} \le {z^{\prime}} \le z_{\xi} + {l \mathord{\left/ {\vphantom {l 2}} \right. \kern-\nulldelimiterspace} 2}$, be the (unknown) surface current. Under the thin wire regime, ${\bf{I}}\left( {{z^{\prime}}} \right)$ is distributed only along the ${\bf{\hat z}}$-axis, i.e., ${\bf{I}}\left( {{z^{\prime}}} \right) \approx {I_z}\left( {{z^{\prime}}} \right) {\bf{\hat z}}$, and it is solution of Pocklington's equation \cite[Eq. (8.22)]{Balanis}:
\begin{equation} \label{Eq_1}
\int\nolimits_{{z_\xi } - l/2}^{{z_\xi } + l/2} {{I_z}\left( {{z^{\prime}}} \right)\left( {\partial _z^2 + k_0^2} \right){G}\left( {z,{z^{\prime}}} \right)d{z^{\prime}}}  =  c_P E_z^{\left( {{\rm{inc}}} \right)}\left( {{\bf r}_\mathcal{S}} \right)
\end{equation}
\noindent where ${G}\left( {z,{z^{\prime}}} \right) = {{\exp \left( { - j{k_0}{R}\left( {z,{z^{\prime}}} \right)} \right)} \mathord{\left/ {\vphantom {{\exp \left( { - j{k_0}r\left( {z,{z^{\prime}}} \right)} \right)} {{R}\left( {z,{z^{\prime}}} \right)}}} \right. \kern-\nulldelimiterspace} {{R}\left( {z,{z^{\prime}}} \right)}}$ is the Green function, ${R}\left( {z,{z^{\prime}}} \right) = \sqrt {{{\left( {z - {z^{\prime}}} \right)}^2} + {a^2}}$, ${c_P} =  - j4\pi \omega {\varepsilon _0}$, and the constraint that ${I_z}\left( {{z^{\prime}}} \right)$ vanishes at the two ends of the antenna, i.e., ${I_z}\left( {{-l/2}} \right) = {I_z}\left( {{l/2}} \right) = 0$ needs to be imposed.

The exact solution of \eqref{Eq_1} usually requires numerical methods, e.g., the method of moments \cite{Balanis}. Under the considered assumptions of minimum scattering radiating elements and thin wire regime, however, the current ${I_z}\left( {{z^{\prime}}} \right)$ in \eqref{Eq_1} can be approximated with the following sinusoidal function \cite{Gradoni}, \cite{Balanis}:
\begin{equation} \label{Eq_2}
{I_z}\left( {{z^{\prime}}} \right) \approx \mathcal{I}\left( {{z_\xi }} \right){{\sin \left( {{k_0}\left( {l/2 - \left| {{z^{\prime}}- z_{\xi}} \right|} \right)} \right)}}/{{\sin \left( {{k_0}l/2} \right)}}
\end{equation}
\noindent where ${z_\xi } - {l \mathord{\left/ {\vphantom {l 2}} \right. \kern-\nulldelimiterspace} 2} \le {z^{\prime}} \le {z_\xi }+  {l \mathord{\left/ {\vphantom {l 2}} \right. \kern-\nulldelimiterspace} 2}$ and $\mathcal{I}\left( {{z_\xi }} \right)$ is the current that flows through the port of the antenna. Based on Sections \ref{Transmitter}-\ref{RIS}, $\mathcal{I}\left( {{z_\xi }} \right)$ is equal to $I_{\rm{Tt}}$, $I_{\rm{Lr}}$, and $I_{\rm{Smn}}$ for the transmitter, receiver, and RIS, respectively. In particular, $\mathcal{I}\left( {{z_\xi }} \right)$ is unknown and is a variable to be estimated. This is discussed next.

The approximation for ${I_z}\left( {{z^{\prime}}} \right)$ in \eqref{Eq_2} assumes that the current density on each thin wire radiating element is not influenced by the proximity of other radiating elements. However, \eqref{Eq_2} can be applied to isolated and closely spaced radiating elements. The accuracy of \eqref{Eq_2} is substantiated in \cite[Sec. III-C]{Gradoni} with the aid of full-wave numerical simulations. In \cite[Fig. 11]{Gradoni}, in particular, it is shown that the approximation in \eqref{Eq_2} is  accurate for typical values of sub-wavelength inter-distances \cite{MDR_JSAC}. We emphasize that the analytical frameworks developed in the following sections rely only on the approximation for ${I_z}\left( {{z^{\prime}}} \right)$ in \eqref{Eq_2} and on the assumption of thin wire regime ($a \ll l$). No additional approximations or assumptions are employed.

\vspace{-0.25cm}
\section{Impedance-Based Mutual Coupling Modeling} \label{MutualCoupling_Modeling} \vspace{-0.10cm}
In this section, we introduce three enabling results: (i) we compute the electric field that is radiated by an isolated antenna; (ii) we introduce the concept of self and mutual impedances; and (iii) we calculate the voltage observed at the ports of a pair of coupled (transmit and receive) antennas.

We consider two arbitrary antennas $\chi  = \left\{ {p,q} \right\}$, which are characterized by the pair $(l_\chi, a_\chi)$ and whose locations are ${\bf{r}}_\chi$ $= x_\chi {\bf{\hat x}} + y_\chi {\bf{\hat y}} + z_\chi {\bf{\hat z}}$. $\chi$ operates in transmit or receive mode, and, in either cases, the current on its surface is ${I_{z,\chi }}\left( {{z^{\prime}}} \right) \approx \mathcal{I}\left( {{z_\chi }} \right)\sin \left( {{k_0}\left( {l_\chi/2 - \left| {{z^{\prime}} - {z_\chi }} \right|} \right)} \right)/\sin \left( {{k_0}l_\chi/2} \right)$ in \eqref{Eq_2}.

\vspace{-0.35cm}
\subsection{Isolated Antenna: Radiated Electric Field} \vspace{-0.10cm}
Based on Section \ref{Assumptions}, we need to compute only the tangential component of the electric field radiated by $\chi$. This field evaluated at the observation point $\bf{r}$ is denoted by $E_{z,\chi }^{\left( {{\rm{rad}}} \right)}\left( {\bf{r}} \right)$.

\begin{lemma}
Consider the transmit antenna $p$. Under the thin wire regime, $E_{z,p }^{\left( {{\rm{rad}}} \right)}\left( {\bf{r}} \right)$ is equal to (${c_E} = - j{\eta _0}/\left( {4\pi {k_0}} \right)$):
\begin{equation} \label{Eq_3}
E_{z,p}^{\left( {{\rm{rad}}} \right)}\left( {\bf{r}} \right) = {c_E}\int\nolimits_{{z_p} - l_p/2}^{{z_p} + l_p/2} {\hspace{-0.25cm}{\mathcal{F}}_p\left( {{\bf{r}},{z^{\prime}}} \right){\mathcal{G}}_p\left( {{\bf{r}},{z^{\prime}}} \right){I_{z,\chi}}\left( {{z^{\prime}}} \right)d{z^{\prime}}}
\end{equation}
\noindent where ${\mathcal{G}}_p\left( {{\bf{r}},{z^{\prime}}} \right) = {\exp \left( { - j{k_0}{\mathcal{R}}_p\left( {{\bf{r}},{z^{\prime}}} \right)} \right)} \mathord{\left/ {\vphantom {{\exp \left( { - j{k_0}{\mathcal{R}}_p\left( {{\bf{r}},{z^{\prime}}} \right)} \right)} {{\mathcal{R}}_p\left( {{\bf{r}},{z^{\prime}}} \right)}}} \right. \kern-\nulldelimiterspace} {{\mathcal{R}}_p\left( {{\bf{r}},{z^{\prime}}} \right)}$ with ${\mathcal{R}}_p\left( {{\bf{r}},{z^{\prime}}} \right) \hspace{-0.05cm} = \hspace{-0.05cm} \sqrt {a_p^2 + {{\left( {z - {z^{\prime}}} \right)}^2}}$ if $\left( {x,y} \right) \hspace{-0.05cm} = \hspace{-0.05cm} \left( {{x_p},{y_p}} \right)$ and ${\mathcal{R}}\left( {{\bf{r}},{z^{\prime}}} \right)$ $= \hspace{-0.1cm} \sqrt {{{\left( {x - {x_p}} \right)}^2} \hspace{-0.02cm} + \hspace{-0.02cm} {{\left( {y - {y_p}} \right)}^2} \hspace{-0.02cm} + \hspace{-0.02cm} {{\left( {z - {z^{\prime}}} \right)}^2}}$ if $\left( {x,y} \right) \hspace{-0.07cm}\ne\hspace{-0.07cm} \left( {{x_p},{y_p}} \right)$, and
\begin{equation} \label{Eq_4}
\begin{split}
{\mathcal{F}}_p\left( {{\bf{r}},{z^{\prime}}} \right) &= \frac{{{{\left( {z - {z^{\prime}}} \right)}^2}}}{{{{\mathcal{R}}_p^2}\left( {{\bf{r}},{z^{\prime}}} \right)}}\left( {\frac{3}{{{{\mathcal{R}}_p^2}\left( {{\bf{r}},{z^{\prime}}} \right)}} + \frac{{3j{k_0}}}{{{\mathcal{R}}_p\left( {{\bf{r}},{z^{\prime}}} \right)}} - k_0^2} \right) \\ 
& - {{\left(j k_0 + {{{{1}} \mathord{\left/
 {\vphantom {{{1}} {{{\mathcal{R}}_p}\left( {{\bf{r}},{z^{\prime}}} \right)}}} \right.
 \kern-\nulldelimiterspace} {{{\mathcal{R}}_p}\left( {{\bf{r}},{z^{\prime}}} \right)}}} \right)} \mathord{\left/
 {\vphantom {{\left( {{{j{k_0}} \mathord{\left/
 {\vphantom {{{1}} {{{\mathcal{R}}_p}\left( {{\bf{r}},{z^{\prime}}} \right)}}} \right.
 \kern-\nulldelimiterspace} {{{\mathcal{R}}_p}\left( {{\bf{r}},{z^{\prime}}} \right)}}} \right)} {{{\mathcal{R}}_p}\left( {{\bf{r}},{z^{\prime}}} \right)}}} \right.
 \kern-\nulldelimiterspace} {{{\mathcal{R}}_p}\left( {{\bf{r}},{z^{\prime}}} \right)}}+ k_0^2
\end{split}
\end{equation}
\begin{IEEEproof}
From Maxwell's equations, the radiated electric field is $E_{z,p }^{\left( {{\rm{rad}}} \right)}\left( {\bf{r}} \right) \hspace{-0.15cm} = \hspace{-0.15cm}  - {c_E}\left( {\partial _z^2 + k_0^2} \right){V_{z,p }}\left( {\bf{r}} \right)$ \cite[Eq. (3.15)]{Balanis}, where ${V_{z,p }}\left( {\bf{r}} \right) \hspace{-0.18cm}  = \hspace{-0.18cm}  \int\nolimits_{{z_p } - {l_p }/2}^{{z_p } + {l_p }/2} {{\mathcal{G}}_p\left( {{\bf{r}},{z^{\prime}}} \right){I_{z,p }}\left( {{z^{\prime}}} \right)d{z^{\prime}}}$ is the normalized magnetic potential, and, under the thin wire regime, ${\mathcal{G}}_p\left( {{\bf{r}},{z^{\prime}}} \right) \hspace{-0.10cm} \approx \hspace{-0.10cm} {\left( {2\pi } \right)^{ - 1}}\int\nolimits_0^{2\pi } {{{\exp \left( { - j{k_0}{\mathcal{R}}_p\left( {{\bf{r}},{z^{\prime}}} \right)} \right)} \mathord{\left/ {\vphantom {{\exp \left( { - j{k_0}{\mathcal{R}}_p\left( {{\bf{r}},{z^{\prime}}} \right)} \right)} {{\mathcal{R}}_p\left( {{\bf{r}},{z^{\prime}}} \right)}}} \right. \kern-\nulldelimiterspace} {{\mathcal{R}}_p\left( {{\bf{r}},{z^{\prime}}} \right)}}d{z^{\prime}}}$.
\end{IEEEproof}
\end{lemma}

\vspace{-0.35cm}
\subsection{Self and Mutual Impedances} \vspace{-0.10cm}
Consider the radiating elements $p$ and $q$, either in transmit or receive mode, and let ${I_{z,p }}\left( {{z^{\prime}}} \right)$ and ${I_{z,q }}\left( {{z^{\prime \prime}}} \right)$ be the currents on the surface of $p$ and $q$, respectively. When $p$ and $q$ are close to each other, whether in transmit or receive mode, some energy created locally within a specific antenna may reach and influence the boundary field of the other antenna. For example, part of the energy radiated from $p$ impinges upon $q$ and excites some currents, which, in turn, generate a scattered field that impinges upon $p$ and excites other currents. This interchange of energy is known as mutual coupling. This effect, which is especially relevant in RISs whose passive scatterers are closely spaced, can be quantified through the mutual independence ${Z_{qp}}$ that characterizes the mutual coupling induced by $p$ on $q$.
\begin{definition}
Let $E_{qp}^{\left( {{\rm{rad}}} \right)}\left( {{z^{\prime \prime}}} \right) = E_{z,p}^{\left( {{\rm{rad}}} \right)}\left( {{{\bf{r}}_{{{\mathcal{S}}_q}}}} \right)$ be the electric field in \eqref{Eq_3} that is radiated by $p$ and is observed on the surface $\mathcal{S}_q$ of $q$, where ${{\bf{r}}_{{{\mathcal{S}}_q}}} = {x_q}{\bf{\hat x}} + {y_q}{\bf{\hat y}} + z^{\prime \prime}{\bf{\hat z}}$ with ${z_q} - {{{l_q}} \mathord{\left/ {\vphantom {{{l_q}} 2}} \right. \kern-\nulldelimiterspace} 2} \le z^{\prime \prime} \le {z_q} + {{{l_q}} \mathord{\left/ {\vphantom {{{l_q}} 2}} \right. \kern-\nulldelimiterspace} 2}$ under the thin wire regime. ${Z_{qp}}$ is defined as:
\begin{equation} \label{Eq_5}
{Z_{qp}} =  - \frac{1}{{\mathcal{I}}\left( {{z_q}} \right){\mathcal{I}}\left( {{z_p}} \right)}\int\nolimits_{{z_q} - {l_q}/2}^{{z_q} + {l_q}/2} {\hspace{-0.15cm} E_{qp}^{\left( {{\rm{rad}}} \right)}\left( {{z^{\prime \prime}}} \right){I_{z,q}}\left( {{z^{\prime \prime}}} \right)d{z^{\prime \prime}}}
\end{equation}
\end{definition}

The self impedance $Z_{pp}$ is obtained from \eqref{Eq_5} by setting $q=p$. By virtue of reciprocity, $Z_{pq} = Z_{qp}$ in vacuum. An important property of ${Z_{qp}} $ is that it depends only on the geometry of the radiating elements, as formalized below.
\begin{lemma}
${Z_{qp}}$ can be explicitly formulated as follows:
\begin{equation} \label{Eq_6}
{Z_{qp}}  = \int\nolimits_{{z_q} - {l_q}/2}^{{z_q} + {l_q}/2} {\hspace{-0.20cm}\int\nolimits_{{z_p} - {l_p}/2}^{{z_p} + {l_p}/2} {\hspace{-0.5cm}{g_{qp}}\left( {{z^{\prime}}}, {{z^{\prime \prime}}} \right){{\tilde I}_{z,p}}\left( {{z^{\prime}}} \right){{\tilde I}_{z,q}}\left( {{z^{\prime \prime}}} \right)d{z^{\prime}}d{z^{\prime \prime}}} }
\end{equation}
\noindent where ${{\tilde I}_{z,\chi }}\left( {{z^{\prime}}} \right) = \sin \left( {{k_0}\left( {{l_\chi }/2 - \left| {{z^{\prime}} - {z_\chi }} \right|} \right)} \right)/\sin \left( {{k_0}{l_\chi }/2} \right)$ and ${g_{qp}}\left( {{z^{\prime}}}, {{z^{\prime \prime}}}\right) = {j{\eta _0}{{\left( {4\pi {k_0}} \right)}^{ - 1}}}  {{\mathcal{F}}_p}\left( {{{\bf{r}}_{{{\mathcal{S}}_q}}},{z^{\prime}}} \right){{\mathcal{G}}_p}\left( {{{\bf{r}}_{{{\mathcal{S}}_q}}},{z^{\prime}}} \right)$.
\begin{IEEEproof}
It follows by inserting \eqref{Eq_3} in \eqref{Eq_5}.
\end{IEEEproof}
\end{lemma}

From \eqref{Eq_6}, we evince that ${Z_{qp}}$ is independent of the (unknown) currents $\mathcal{I}\left( {{z_\chi }} \right)$ that flow through the ports of $\chi  = \left\{ {p,q} \right\}$. This confirms that ${Z_{qp}}$ depends only on the geometry of the antennas and is independent of the voltage generators and the (tunable) loads. Thus, it can be precomputed once and then used for different RIS-assisted wireless systems. The usefulness of ${Z_{qp}}$ will become apparent in further text.

\vspace{-0.35cm}
\subsection{Pair of Antennas: Two Transmit Antennas} \vspace{-0.10cm}
Consider the antenna elements $p$ and $q$, and assume that they both operate in transmit mode. Thus, they are fed by two generators $V_{\rm{Gp}}$ and $V_{\rm{Gq}}$ as described in Section \ref{Transmitter}. The objective of this section is to compute an explicit analytical relation between the voltages $V_{\rm{Tp}}$ and $V_{\rm{Tq}}$ and the currents $I_{\rm{Tp}}=\mathcal{I}\left( {{z_p }} \right)$ and $I_{\rm{Tq}}=\mathcal{I}\left( {{z_q }} \right)$ as a function of the mutual coupling between $p$ and $q$. This is possible by using the self and mutual impedances ${Z_{qp}}$, as stated in the following lemma.
\begin{lemma} \label{Lemma_VoltagesCurrents}
Consider two transmit antennas $p$ and $q$. The voltages and currents at their ports are interwoven through the following linear system (known as constitutive equations):
\begin{equation} \label{Eq_7}
\left\{ \begin{array}{l}
{V_{{\rm{Tp}}}} = {Z_{pp}}{I_{{\rm{Tp}}}} + {Z_{pq}}{I_{{\rm{Tq}}}}\\
{V_{{\rm{Tq}}}} = {Z_{qp}}{I_{{\rm{Tp}}}} + {Z_{qq}}{I_{{\rm{Tq}}}}
\end{array} \right.
\end{equation}
\noindent where ${Z_{qp}}$ is the mutual impedance in \eqref{Eq_6}.
\begin{IEEEproof}
From Maxwell's equations, the radiated electric field on the surface $\mathcal{S}_q$ of $q$, i.e., evaluated at ${{\bf{r}}_{{{\mathcal{S}}_q}}} = {x_q}{\bf{\hat x}} + {y_q}{\bf{\hat y}} + z^{\prime \prime}{\bf{\hat z}}$ with ${z_q} - {{{l_q}} \mathord{\left/ {\vphantom {{{l_q}} 2}} \right. \kern-\nulldelimiterspace} 2} \le z^{\prime \prime} \le {z_q} + {{{l_q}} \mathord{\left/ {\vphantom {{{l_q}} 2}} \right. \kern-\nulldelimiterspace} 2}$ under the thin wire regime, is equal to $E_{z,q}^{\left( {{\rm{rad}}{\rm{, mc}}} \right)}\left( {{{\bf{r}}_{{{\mathcal{S}}_q}}}} \right) = E_{z,p}^{\left( {{\rm{rad}}} \right)}\left( {{{\bf{r}}_{{{\mathcal{S}}_q}}}} \right) + E_{z,q}^{\left( {{\rm{rad}}} \right)}\left( {{{\bf{r}}_{{{\mathcal{S}}_q}}}} \right)$ in the presence of $p$, where ``mc'' stands for mutual coupling and the two addends can be computed from \eqref{Eq_3}. By definition of delta-gap model, the voltage generator is modeled as an incident electric field that is non-zero just outside the surface of the gap, while the gap itself is filled with a perfectly conducting material. Therefore, the incident electric field on $\mathcal{S}_q$ is $E_{z,q}^{\left( {{\rm{inc}}} \right)}\left( {{{\bf{r}}_{{{\mathcal{S}}_q}}}} \right) = {V_{{\rm{Tq}}}}\delta \left( {{z^{\prime \prime} - z_q}} \right)$, and the boundary conditions for perfectly conducting wires apply to the entire surface $\mathcal{S}_q$ of $q$. For perfectly conducting wires, in particular, the boundary conditions on $\mathcal{S}_q$ impose $E_{z,q}^{\left( {{\rm{inc}}} \right)}\left( {{{\bf{r}}_{{{\mathcal{S}}_q}}}} \right) + E_{z,q}^{\left( {{\rm{rad}}{\rm{, mc}}} \right)}\left( {{{\bf{r}}_{{{\mathcal{S}}_q}}}} \right) = 0$, which yields $E_{z,p}^{\left( {{\rm{rad}}} \right)}\left( {{{\bf{r}}_{{{\mathcal{S}}_q}}}} \right)$ $+ E_{z,q}^{\left( {{\rm{rad}}} \right)}\left( {{{\bf{r}}_{{{\mathcal{S}}_q}}}} \right) =  - {V_{{\rm{Tq}}}}\delta \left( {{z^{\prime \prime} -z_q}} \right)$. Consider the integral:
\begin{equation} \label{Eq_8}
{{\mathcal{J}}_q} =  - \frac{1}{{{{\mathcal{I}}_{z,q}}\left( {{z_q}} \right)}}\int\nolimits_{{z_q} - {l_q}/2}^{{z_q} + {l_q}/2} {E_{z,q}^{\left( {{\rm{rad}}{\rm{, mc}}} \right)}\left( {{{\bf{r}}_{{{\mathcal{S}}_q}}}} \right){I_{z,q}}\left( {{z^{\prime \prime}}} \right)d{z^{\prime \prime}}}
\end{equation}
From \eqref{Eq_5}, by definition, we obtain ${{\mathcal{J}}_q} = {Z_{qp}}{I_{{\rm{Tp}}}} + {Z_{qq}}{I_{{\rm{Tq}}}}$. From the boundary conditions, we obtain ${{\mathcal{J}}_q} =  - \frac{1}{{{{\mathcal{I}}_{z,q}}\left( {{z_q}} \right)}}\int\nolimits_{{z_q} - {l_q}/2}^{{z_q} + {l_q}/2} {\left( { - {V_{{\rm{Tq}}}}\delta \left( {{z^{\prime \prime}} - {z_q}} \right)} \right){I_{z,q}}\left( {{z^{\prime \prime}}} \right)d{z^{\prime \prime}}}  = {V_{{\rm{Tq}}}}$. By comparing these two results, the second equation in \eqref{Eq_7} is proved. The first equation follows \textit{mutatis mutandis}.
\end{IEEEproof}
\end{lemma}

\vspace{-0.35cm}
\subsection{Pair of Antennas: One Transmit and One Receive Antenna} \vspace{-0.10cm}
Consider the antenna elements $p$ and $q$ that operate in transmit and receive mode, respectively. Therefore, $p$ is driven by a voltage generator according to Section \ref{Transmitter} and the port of $q$ is connected to a load impedance according to Sections \ref{RIS} and \ref{Receiver}. The relation between the voltages $V_{\rm{Tp}}$ and $V_{\rm{Lq}}$ or $V_{\rm{Smn}}$ and the currents $I_{\rm{Tp}}=\mathcal{I}\left( {{z_p }} \right)$ and $I_{\rm{Lq}}=\mathcal{I}\left( {{z_q }} \right)$ or $I_{\rm{Smn}}=\mathcal{I}\left( {{z_q }} \right)$ at the ports of $p$ and $q$ is given as follows.
\begin{lemma} \label{Lemma_VoltagesCurrents_pq}
Consider one transmit antenna $p$ and one receive antenna $q$. The voltages and currents at their ports are interwoven through the following system of linear equations:
\begin{equation} \label{Eq_9}
\left\{ \begin{array}{l}
{V_{{\rm{Tp}}}} = {Z_{pp}}{I_{{\rm{Tp}}}} + {Z_{pq}}{I_{{\rm{q}}}}\\
{V_{{\rm{q}}}} = {Z_{qp}}{I_{{\rm{Tp}}}} + {Z_{qq}}{I_{{\rm{q}}}}
\end{array} \right.
\end{equation}
\noindent where $V_{\rm{q}} = V_{\rm{Lq}}$ and $I_{\rm{q}} = I_{\rm{Lq}}$ for the receiver, $V_{\rm{q}} = V_{\rm{Smn}}$ and $I_{\rm{q}} = I_{\rm{Smn}}$ for the RIS, and ${Z_{qp}}$ is given in \eqref{Eq_6}.
\begin{IEEEproof}
It is similar to the proof of Lemma \ref{Lemma_VoltagesCurrents} with the main difference that $q$ operates in receive mode, i.e., $q$ is not driven by a voltage generator but is connected to a load impedance. The voltage drop across the load impedance can be viewed as generated by an incident electric field that is non-zero just outside the surface of the gap occupied by the load, while the gap itself is filled with a perfectly conducting material. By definition, the equivalent incident electric field is $E_{z,q}^{\left( {{\rm{inc}}} \right)}\left( {{{\bf{r}}_{{{\rm{S}}_q}}}} \right) = {V_{{\rm{q}}}}\delta \left( {{z^{\prime \prime} - z_q}} \right)$ with ${V_{{\rm{q}}}}$ given in Sections \ref{Receiver} and \ref{RIS}. The rest of the proof is the same as Lemma \ref{Lemma_VoltagesCurrents}. 
\end{IEEEproof}
\end{lemma}

\vspace{-0.25cm}
\section{End-to-End Communication Model} \label{E2E_Model} \vspace{-0.10cm}
In this section, based on the enabling results proved in Lemmas \ref{Lemma_VoltagesCurrents} and \ref{Lemma_VoltagesCurrents_pq}, we introduce an end-to-end communication model for RIS-assisted communications that accounts for an arbitrary number of coupled antenna elements at the transmitter and receiver, and passive scatterers at the RIS. Some special system setups are analyzed to gain engineering insights.

\vspace{-0.35cm}
\subsection{End-to-End Equivalent Channel Model} \vspace{-0.10cm}
Let ${{\bf{V}}_{\rm{G}}}$ be the $N_{\rm{t}} \times 1$ vector whose $t$th entry is $V_{\rm{Tt}}$ and ${{\bf{V}}_{\rm{L}}}$ be the $N_{\rm{r}} \times 1$ vector whose $r$th entry is $V_{\rm{Lr}}$. The term \textit{end-to-end} communication model is referred to an analytical framework that formulates ${{\bf{V}}_{\rm{L}}}$ as a function of ${{\bf{V}}_{\rm{G}}}$. More precisely, we wish to find an $N_{\rm{r}} \times N_{\rm{t}}$ matrix, ${{\bf{\mathbfcal{H}}}_{{\rm{E2E}}}}$, such that ${{\bf{V}}_{\rm{L}}} = {{{\mathbfcal{H}}}_{{\rm{E2E}}}} {{\bf{V}}_{\rm{G}}}$. The matrix ${{\bf{\mathbfcal{H}}}_{{\rm{E2E}}}}$, which is referred to as end-to-end equivalent communication channel, accounts for the transmit and receive antenna elements, the passive scatterers of the RIS, the load impedances at the receiver, and, more importantly, the tunable load impedances of the RIS. The matrix ${{\bf{\mathbfcal{H}}}_{{\rm{E2E}}}}$ is given in the following theorem.
\begin{theorem} \label{Proposition_HE2E}
Let ${{\mathbfcal{Z}}_{{\rm{XY}}}}$ be the ${\rm{N_x}} \times {\rm{N_y}}$ matrix whose $({\rm x}, {\rm y})$th entry is the mutual impedance in \eqref{Eq_6} between the $\rm x$th and $\rm y$th radiating elements of $\rm X$ and $\rm Y$, where ${\rm{X, Y}} = \left\{ {{\rm{T}},{\rm{S}},{\rm{R}}} \right\}$ and ${{\rm{N_x}}}, {{\rm{N_y}}} = \left\{ {{{\rm{N}}_{\rm{t}}},{{\rm{N}}_{{\rm{ris}}}},{{\rm{N}}_{\rm{r}}}} \right\}$, and $\rm T$,  $\rm S$,  $\rm R$ identify the transmitter, the RIS, and the receiver, respectively. ${{\bf{\mathbfcal{H}}}_{{\rm{E2E}}}}$ is as follows:
\begin{eqnarray} \label{Eq_10}
{{\mathbfcal{H}}_{{\rm{E2E}}}}\hspace{-0.3cm} &  = & \hspace{-0.3cm} {\left( {{{\bf{I}}_{N_{\rm{r}} \times N_{\rm{r}}}} + {{\mathbfcal{P }}_{{\rm{RSR}}}}{\bf{Z}}_{\rm{L}}^{ - 1} - {{\mathbfcal{P }}_{{\rm{RST}}}}{\mathbfcal{P }}_{{\rm{GTST}}}^{ - 1}{{\mathbfcal{P }}_{{\rm{TSR}}}}{\bf{Z}}_{\rm{L}}^{ - 1}} \right)^{ - 1}} \nonumber \\ \hspace{-0.3cm} &  \times & \hspace{-0.3cm} {{\mathbfcal{P }}_{{\rm{RST}}}}{\mathbfcal{P }}_{{\rm{GTST}}}^{ - 1}
\end{eqnarray}
\noindent where ${{\bf{I}}_{N_{\rm{r}} \times N_{\rm{r}}}}$ denotes an $N_{\rm{r}} \times N_{\rm{r}}$  identity matrix, ${{\bf{Z}}_{\rm{G}}}$ is the $N_{\rm{t}} \times N_{\rm{t}}$ diagonal matrix whose $(t,t)$th entry is ${{{Z}}_{\rm{Gt}}}$, ${{\bf{Z}}_{\rm{RIS}}}$ is the $N_{\rm{ris}} \times N_{\rm{ris}}$ diagonal matrix whose $(mn,mn)$th entry is ${{{Z}}_{\rm{Smn}}}$, ${{\bf{Z}}_{\rm{L}}}$ is the $N_{\rm{r}} \times N_{\rm{r}}$ diagonal matrix whose $(r,r)$th entry is ${{{Z}}_{\rm{Lr}}}$, ${{\mathbfcal{P }}_{{\rm{GTST}}}} = {{\bf{Z}}_{\rm{G}}} + {{\mathbfcal{P }}_{{\rm{TST}}}}$, and:
\begin{equation} \label{Eq_11}
{{\mathbfcal{P}}_{{\rm{XSY}}}} = {{\mathbfcal{Z}}_{{\rm{XY}}}} - {{\mathbfcal{Z}}_{{\rm{XS}}}}{\left( {{{\bf{Z}}_{{\rm{RIS}}}} + {{\mathbfcal{Z}}_{{\rm{SS}}}}} \right)^{ - 1}}{{\mathbfcal{Z}}_{{\rm{SY}}}}
\end{equation}
\begin{IEEEproof}
Let ${{\bf{V}}_{\rm{T}}}$ and ${{\bf{I}}_{\rm{T}}}$ be the $N_{\rm{t}} \hspace{-0.10cm}\times \hspace{-0.10cm} 1$ vectors whose $t$th entries are $V_{\rm{Tt}}$ and $I_{\rm{Tt}}$, so that ${{\bf{V}}_{\rm{T}}} \hspace{-0.10cm}=\hspace{-0.10cm} {{\bf{V}}_{\rm{G}}} \hspace{-0.05cm}-\hspace{-0.05cm} {{\bf{Z}}_{\rm{G}}}{{\bf{I}}_{\rm{T}}}$; ${{\bf{I}}_{\rm{L}}}$ be the $N_{\rm{r}} \hspace{-0.05cm} \times \hspace{-0.05cm} 1$ vector whose $r$th entry is $I_{\rm{Lr}}$ so that ${{\bf{V}}_{\rm{L}}} \hspace{-0.10cm}= \hspace{-0.10cm} - {{\bf{Z}}_{\rm{L}}}{{\bf{I}}_{\rm{L}}}$; and ${{\bf{V}}_{\rm{S}}}$ and ${{\bf{I}}_{\rm{S}}}$ be the $N_{\rm{ris}} \hspace{-0.05cm} \times \hspace{-0.05cm} 1$ vectors whose $mn$th entries are $V_{\rm{Smn}}$ and $I_{\rm{Smn}}$, so that ${{\bf{V}}_{\rm{S}}} \hspace{-0.10cm}= \hspace{-0.10cm}- {{\bf{Z}}_{\rm{RIS}}}{{\bf{I}}_{\rm{S}}}$. The proof is obtained in two steps. (i) The proofs of Lemmas \ref{Lemma_VoltagesCurrents} and \ref{Lemma_VoltagesCurrents_pq} are applied to the system with $N_{\rm{t}}$ transmit antennas, $N_{\rm{r}}$ receive antennas, and $N_{\rm{ris}}$ passive scatterers. This yields ${{\bf{V}}} \hspace{-0.10cm} = \hspace{-0.10cm} {{\mathbfcal{Z}}}{{\bf{I}}}$, where ${{\bf{V}}} \hspace{-0.10cm} = \hspace{-0.10cm} \left[ {{{\bf{V}}_{\rm{T}}},{{\bf{V}}_{\rm{S}}},{{\bf{V}}_{\rm{L}}}} \right]$, ${{\bf{I}}} \hspace{-0.10cm} = \hspace{-0.10cm} \left[ {{{\bf{I}}_{\rm{T}}},{{\bf{I}}_{\rm{S}}},{{\bf{I}}_{\rm{L}}}} \right]$, and ${{\mathbfcal{Z}}} \hspace{-0.10cm} = \hspace{-0.10cm} \left[ {{{\mathbfcal{Z}}_{{\rm{TT}}}},{{\mathbfcal{Z}}_{{\rm{TS}}}},{{\mathbfcal{Z}}_{{\rm{TR}}}};{{\mathbfcal{Z}}_{{\rm{ST}}}},{{\mathbfcal{Z}}_{{\rm{SS}}}},{{\mathbfcal{Z}}_{{\rm{SR}}}};{{\mathbfcal{Z}}_{{\rm{RT}}}},{{\mathbfcal{Z}}_{{\rm{RS}}}},{{\mathbfcal{Z}}_{{\rm{RR}}}}} \right]$ are block matrices. (ii) The system of equations ${{\bf{V}}}\hspace{-0.10cm} = \hspace{-0.10cm} {{\mathbfcal{Z}}}{{\bf{I}}}$ is solved using ${{\bf{V}}_{\rm{T}}} \hspace{-0.10cm} = \hspace{-0.10cm} {{\bf{V}}_{\rm{G}}} - {{\bf{Z}}_{\rm{G}}}{{\bf{I}}_{\rm{T}}}$, ${{\bf{V}}_{\rm{L}}}\hspace{-0.10cm}  = \hspace{-0.10cm}  - {{\bf{Z}}_{\rm{L}}}{{\bf{I}}_{\rm{L}}}$, and ${{\bf{V}}_{\rm{S}}}\hspace{-0.10cm}  = \hspace{-0.10cm} - {{\bf{Z}}_{\rm{RIS}}}{{\bf{I}}_{\rm{S}}}$. 
\end{IEEEproof}
\end{theorem}

The matrix ${{\bf{\mathbfcal{H}}}_{{\rm{E2E}}}}$ is an EM-compliant end-to-end multiple-input multiple-output communication channel model that can be utilized to quantify the advantages and limitations of RISs. By computing, e.g., the rank, the eigenvalues, and the eigenvectors of ${{\bf{\mathbfcal{H}}}_{{\rm{E2E}}}}$, the multiplexing vs. diversity tradeoff of RIS-assisted systems can be unveiled. In particular, the matrix of tunable impedances ${{\bf{Z}}_{\rm{RIS}}}$ can be appropriately optimized for realizing the desired multiplexing vs. diversity tradeoff.

\vspace{-0.25cm}
\subsection{Design Insights} \vspace{-0.10cm}
${{\bf{\mathbfcal{H}}}_{{\rm{E2E}}}}$ accounts for the mutual coupling among the $\left(N_{\rm{r}}+N_{\rm{ris}}+N_{\rm{t}}\right)$ available radiating elements. To substantiate the consistency of ${{\bf{\mathbfcal{H}}}_{{\rm{E2E}}}}$ and to gain engineering insights for system design, we analyze the following relevant case study.
\begin{corollary} \label{Corollary_E2E}
Consider (i) $N_{\rm{t}} = N_{\rm{r}} = 1$ and (ii) large transmission distances between the transmitter and the receiver, the transmitter and the RIS, and the RIS and the receiver. Then, ${{\bf{\mathcal{H}}}_{{\rm{E2E}}}}$ in \eqref{Eq_10} is a scalar and simplifies as follows:
\begin{equation} \label{Eq_12}
{{\mathcal{H}}_{{\rm{E2E}}}} = {{\mathcal{Y}}_0}\left( {{{\mathcal{Z}}_{{\rm{RT}}}} - {{\mathbfcal{Z}}_{{\rm{RS}}}}{{\left( {{{\bf{Z}}_{{\rm{RIS}}}} + {{\mathbfcal{Z}}_{{\rm{SS}}}}} \right)}^{ - 1}}{{\mathbfcal{Z}}_{{\rm{ST}}}}} \right)
\end{equation}
\noindent where ${{\mathcal{Y}}_0} = {{\rm{Z}}_{\rm{L}}}{\left( {{{\rm{Z}}_{\rm{L}}} + {{\mathcal{Z}}_{{\rm{RR}}}}} \right)^{ - 1}}{\left( {{{\rm{Z}}_{\rm{G}}} + {{\mathcal{Z}}_{{\rm{TT}}}}} \right)^{ - 1}}$, and ${{{\rm{Z}}_{\rm{G}}}}$, ${{{\rm{Z}}_{\rm{L}}}}$, ${{{\mathcal{Z}}_{\rm{TT}}}}$, ${{{\mathcal{Z}}_{\rm{RR}}}}$, ${{{\mathcal{Z}}_{\rm{RT}}}}$ are scalar versions of the associated matrices.
\begin{IEEEproof}
In the large distance regime, based on \eqref{Eq_3}, the absolute value of $Z_{qp}$ in \eqref{Eq_6} for $p \ne q$ decays linearly with the radial distance between the radiating elements. On the other hand, $Z_{pp}$ is independent of the radial distance. Then, from \eqref{Eq_11}, we have ${{\mathbfcal{P}}_{{\rm{TST}}}} \approx {{\mathbfcal{Z}}_{{\rm{TT}}}}$ and ${{\mathbfcal{P}}_{{\rm{RSR}}}} \approx {{\mathbfcal{Z}}_{{\rm{RR}}}}$. Also, ${\rm{1}} + {{\mathbfcal{Z}}_{{\rm{RR}}}}{\bf{Z}}_{\rm{L}}^{ - 1} - {\left( {{{\bf{Z}}_{\rm{G}}} + {{\mathbfcal{Z}}_{{\rm{TT}}}}} \right)^{ - 1}}{\bf{Z}}_{\rm{L}}^{ - 1}{{\mathbfcal{P}}_{{\rm{RST}}}}{{\mathbfcal{P}}_{{\rm{TSR}}}} \approx {\rm{1}} + {{\mathbfcal{Z}}_{{\rm{RR}}}}{\bf{Z}}_{\rm{L}}^{ - 1}$. The proof follows from \eqref{Eq_10}, by setting $N_{\rm{t}} = N_{\rm{r}} = 1$.
\end{IEEEproof}
\end{corollary}

From Corollary \ref{Corollary_E2E}, the following remarks can be made.

\noindent \textbf{\textit{Far-Field Path-Loss}}. ${{\bf{\mathcal{H}}}_{{\rm{E2E}}}}$ in \eqref{Eq_12} is the sum of the line-of-sight (LOS) link ${{\bf{\mathcal{H}}}_{{\rm{LOS}}}} \hspace{-0.10cm} = \hspace{-0.10cm} {{{\rm{Z}}_{{\rm{RT}}}}}$ and the \textit{virtual} LOS link ${{\bf{\mathcal{H}}}_{{\rm{VLOS}}}} \hspace{-0.1cm} = \hspace{-0.1cm} {{\mathbfcal{Z}}_{{\rm{RS}}}}{{\left( {{{\bf{Z}}_{{\rm{RIS}}}} + {{\mathbfcal{Z}}_{{\rm{SS}}}}} \right)}^{ - 1}}{{\mathbfcal{Z}}_{{\rm{ST}}}}$. Since the received power is proportional to ${P_{\rm{L}}} \hspace{-0.1cm} \propto \hspace{-0.1cm} {\left| {{{\mathcal{H}}_{{\rm{E2E}}}}} \right|^2}$, we retrieve the expected scaling laws as a function of the transmitter-receiver distance ($r_{\rm{RT}}$), i.e., $\left| {{{\mathcal{H}}}_{{\rm{LOS}}}} \right|^2 \hspace{-0.05cm} \propto \hspace{-0.05cm} r_{\rm{RT}}^{-2}$, and the transmitter-RIS ($r_{\rm{ST}}$) and RIS-receiver ($r_{\rm{RS}}$) distances, i.e., $\left|{{{\mathcal{H}}}_{{\rm{VLOS}}}}\right|^2 \hspace{-0.05cm} \propto \hspace{-0.05cm} r_{\rm{RS}}^{-2} r_{\rm{ST}}^{-2}$ \cite{MDR_PathLoss}.

\noindent \textbf{\textit{Reconfigurability of the RIS}}. ${{\bf{\mathcal{H}}}_{{\rm{E2E}}}}$ in \eqref{Eq_12} explicitly depends on the tunable load impedances ${{\bf{Z}}_{{\rm{RIS}}}}$, which can be appropriately optimized for obtaining the desired performance. It is worth noting that the intertwinement between the amplitude and phase response of the individual passive scatterers of the RIS is inherently accounted for in ${{\bf{\mathcal{H}}}_{{\rm{E2E}}}}$ through ${{\bf{Z}}_{{\rm{RIS}}}}$ \cite{RuiZhang}, and it depends on the circuital model of the tuning circuit.

\noindent \textbf{\textit{Conventional vs. Mutual Coupling Modeling}}. ${{\bf{\mathcal{H}}}_{{\rm{E2E}}}}$ in \eqref{Eq_12} explicitly accounts for the mutual coupling among the passive scatterers of the RIS through the mutual impedances ${{\mathbfcal{Z}}_{{\rm{SS}}}}$. In particular, ${{\mathcal{H}}_{{\rm{VLOS}}}} \hspace{-0.1cm} = \hspace{-0.1cm} \sum\nolimits_{u = 1}^{{N_{{\rm{ris}}}}} {\sum\nolimits_{v = 1}^{{N_{{\rm{ris}}}}} {{\mathbf{\Phi}}\left( {v,u} \right){{\mathbfcal{Z}}_{{\rm{ST}}}}\left( u \right){{\mathbfcal{Z}}_{{\rm{RS}}}}\left( v \right)} }$, where ${{{\mathbfcal{Z}}_{{\rm{ST}}}}\left( u \right)}$ and ${{{\mathbfcal{Z}}_{{\rm{RS}}}}\left( v \right)}$ are the $u$th and $v$th entries of ${{{\mathbfcal{Z}}_{{\rm{ST}}}}}$ and ${{{\mathbfcal{Z}}_{{\rm{RS}}}}}$, respectively, and ${{\mathbf{\Phi}}\left( {v,u} \right)}$ is the $(v,u)$th entry of ${\mathbf{\Phi}}  \hspace{-0.1cm} =  \hspace{-0.1cm} {\left( {{{\bf{Z}}_{{\rm{RIS}}}} + {{\mathbfcal{Z}}_{{\rm{SS}}}}} \right)^{ - 1}}$. In the absence of mutual coupling, ${{\mathbfcal{Z}}_{{\rm{SS}}}}$ is a diagonal matrix and ${{\mathcal{H}}_{{\rm{VLOS}}}}$ simplifies to ${\mathcal{H}}_{{\rm{VLOS}}}^{\left( {{\rm{no \, coupling}}} \right)} \hspace{-0.1cm} = \hspace{-0.1cm} \sum\nolimits_{u = 1}^{{N_{{\rm{ris}}}}} {{\mathbf{\Phi}}\left( {u,u} \right){{\mathbfcal{Z}}_{{\rm{ST}}}}\left( u \right){{\mathbfcal{Z}}_{{\rm{RS}}}}\left( u \right)}$. The obtained ${{\bf{\mathcal{H}}}_{{\rm{E2E}}}}$ subsumes current frameworks that are mutual coupling unaware, but it allows for the optimization of the tunable loads of the RIS by accounting for their mutual coupling.

\noindent \textbf{\textit{How to Utilize the Proposed Communication Model?}} ${\mathcal{H}}_{{\rm{VLOS}}}^{\left( {{\rm{no \, coupling}}} \right)}$ \textit{resembles} communication-theoretic models that are formulated in terms of the received signal-to-noise ratio (SNR) and that are typically used in wireless communications \cite[Eq. (1)]{Xuewen}. In particular, ${{\mathbf{\Phi}}\left( {u,u} \right)}$ can be viewed as the reflection coefficient of the $u$th passive scatterer (unit cell) of the RIS. ${{\mathbf{\Phi}}\left( {v,u} \right)}$ in ${\mathcal{H}}_{{\rm{VLOS}}}$ has a similar meaning, but it accounts for the mutual coupling between the unit cells $u$ and $v$. The main differences with SNR-based models are: (i) ${\mathcal{H}}_{{\rm{VLOS}}}$ is based on mutual impedances; (ii) ${\mathcal{H}}_{{\rm{VLOS}}}$ yields a one-to-one mapping among the voltages and the currents that can be \textit{measured} at the ports of the transmitter and receiver; and (iii) ${\mathcal{H}}_{{\rm{VLOS}}}$ stems directly from Maxwell's equations. The \textit{analytical} similarity with widely used EM-unaware communication models allows us to use ${\mathcal{H}}_{{\rm{VLOS}}}$ for system optimization by using algorithms similar to those used for the SNR, with the advantage that the proposed model is end-to-end, EM-compliant, unit cell aware, and mutual coupling aware. A tangible example of how to utilize the proposed model for optimizing the load impedances (i.e., the tunable circuits) of the RIS can be found in \cite{Xuewen_OptRIS}.

\begin{figure}[!t]
\centering
\includegraphics[width=0.78\columnwidth]{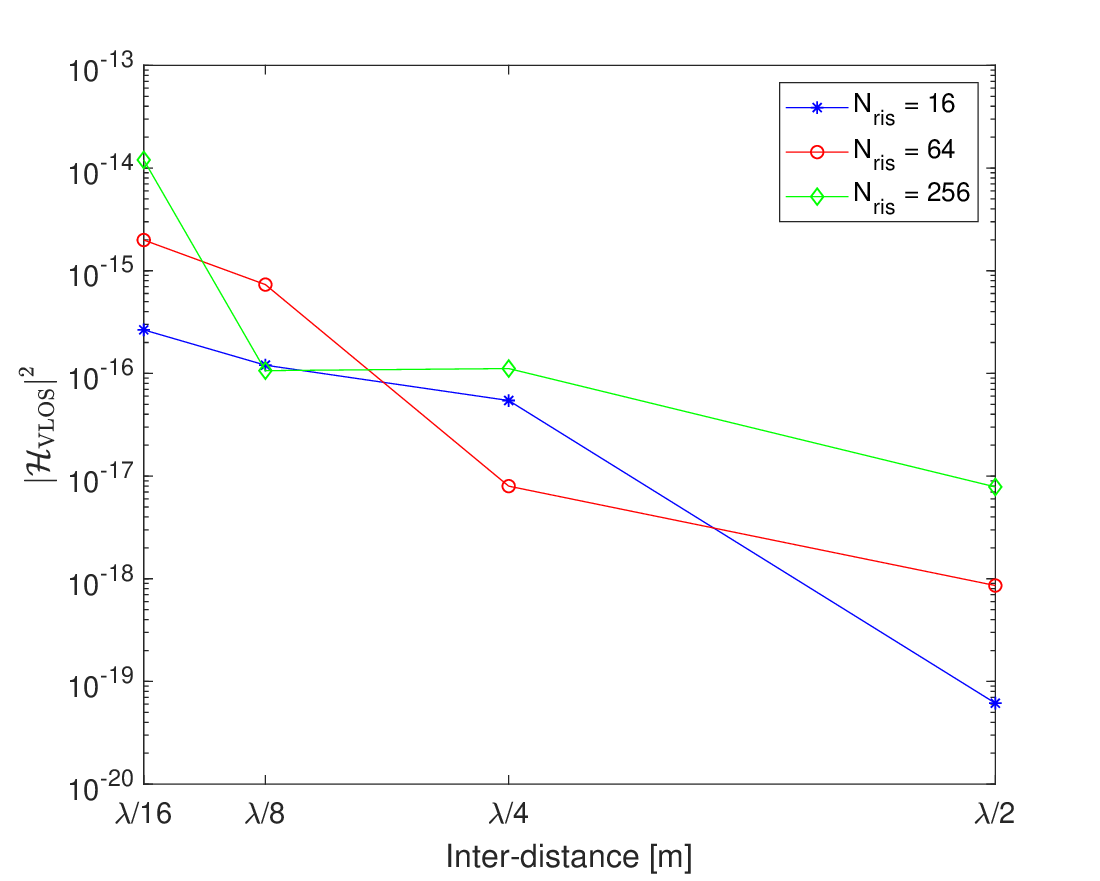}
\vspace{-0.1cm} \caption{Impact of mutual coupling in RIS-assisted transmission.}
\label{Fig_2} \vspace{-0.30cm}
\end{figure}
\vspace{-0.25cm}
\section{Numerical Results} \label{Numerical_Results} \vspace{-0.10cm}
In this section, we report some numerical results in order to study the impact of mutual coupling. In Fig. \ref{Fig_2}, we illustrate $\left|{{\bf{\mathcal{H}}}_{{\rm{VLOS}}}}\right|^2 = \left|{{\mathbfcal{Z}}_{{\rm{RS}}}}{{\left( {{{\bf{Z}}_{{\rm{RIS}}}} + {{\mathbfcal{Z}}_{{\rm{SS}}}}} \right)}^{ - 1}}{{\mathbfcal{Z}}_{{\rm{ST}}}}\right|^2$ in \eqref{Eq_12} for $\mathcal{Y}_0 = 1$ as a function of the inter-distance $d$ and the number $N_{\rm{ris}}$ of scattering elements of the RIS. The setup is the following: $f= 28$ GHz, $N_{\rm{t}} = N_{\rm{r}}=1$, ${{\bf{r}}_{t}} = (5, -5, 3)$, ${{\bf{r}}_{r}} = (5, 5, 1)$, the RIS is centered at $(0,0,0)$ with $M=N=\sqrt{N_{\rm{ris}}}$, the transmit and receive antennas and the passive scatterers of the RIS are identical with $a=\lambda/500$, $l=\lambda/32$, and ${Z_{{\rm{Smn}}}} = {R_{{\rm{Smn}}}} + j\omega {L_{{\rm{Smn}}}}$ with ${R_{{\rm{Smn}}}} = 1$ $\Omega$ and ${L_{{\rm{Smn}}}} = 1$ nH.

In the considered setup, we observe that the mutual coupling among the elements of the RIS has a noticeable impact on the received intensity ${\left| {{{\bf{\mathcal{H}}}_{{\rm{VLOS}}}}} \right|^2}$. The results in Fig. \ref{Fig_2} substantiate the need of mutual coupling aware designs, in which the tunable loads $Z_{\rm{Smn}}$ are optimized to maximize ${\left| {{{\bf{\mathcal{H}}}_{{\rm{VLOS}}}}} \right|^2}$ \cite{Xuewen_OptRIS}.

\vspace{-0.15cm}
\section{Conclusion} \label{Conclusion} \vspace{-0.10cm}
We have introduced an end-to-end, EM-compliant, mutual coupling aware, and unit cell aware communication model for RIS-assisted wireless systems. The proposed model is inherently compatible with conventional communication-theoretic frameworks and can be leveraged for the physics-compliant modeling, analysis, and optimization of RIS-assisted communications. The validation of the proposed model with the aid of hardware platforms and experimental measurements is a natural and relevant continuation of the present research work.

\bibliographystyle{IEEEtran}

\end{document}